\documentclass[prl,twocolumn,showpacs,superscriptaddress,floatfix]{revtex4}

\usepackage{epsfig}
\usepackage{graphics}  
\usepackage{makeidx}
\usepackage{colordvi}
\usepackage{amsmath}
\RequirePackage{xspace}
\usepackage{relsize}

\def\pep2{PEP-II \xspace}

\def\babar{\mbox{\slshape B\kern-0.1em{\smaller A}\kern-0.1em
    B\kern-0.1em{\smaller A\kern-0.2em R}}\xspace}

\def\invfb   {\ensuremath{\mbox{\,fb}^{-1}}\xspace}
\def\epem       {\ensuremath{e^+e^-}\xspace}

\def\CP                {\ensuremath{C\!P}\xspace}

\def\qqbar {\ensuremath{q\overline q}\xspace}

\def\ccbar {\ensuremath{c\overline c}\xspace}

\def\Y#1S{\ensuremath{\Upsilon{(#1S)}}\xspace}

\def\Bbar    {\kern 0.18em\overline{\kern -0.18em B}{}\xspace}

\def\BB      {\ensuremath{B\Bbar}\xspace} 

\def\Kbar  {\kern 0.2em\overline{\kern -0.2em K}{}\xspace}

\def\Kz    {\ensuremath{K^0}\xspace}
\def\Kzb   {\ensuremath{\Kbar^0}\xspace}
\def\KzKzb {\ensuremath{\Kz \kern -0.16em \Kzb}\xspace}
\def\KS    {\ensuremath{K^0_{\scriptscriptstyle S}}\xspace}

\newcommand{\gevc}{\ensuremath{{\mathrm{\,Ge\kern -0.1em V\!/}c}}\xspace}
\newcommand{\mevc}{\ensuremath{{\mathrm{\,Me\kern -0.1em V\!/}c}}\xspace}
\newcommand{\gevcc}{\ensuremath{{\mathrm{\,Ge\kern -0.1em V\!/}c^2}}\xspace}
\newcommand{\mevcc}{\ensuremath{{\mathrm{\,Me\kern -0.1em V\!/}c^2}}\xspace}

\newcommand{\stat}{\ensuremath{\mathrm{(stat)}}\xspace}
\newcommand{\syst}{\ensuremath{\mathrm{(syst)}}\xspace}

\newcommand{\progtp}    [1]  {{Prog.\ Theor.\ Phys.\ {\bf #1}}}

\begin{document}  

\begin{flushleft}
arXiv:1011.5477 (hep-ex)\\
\babar-PUB-10/027\\
SLAC-PUB-14314\\
\end{flushleft}

\title{
{\large  \boldmath
Search for \CP Violation in the Decay $D^\pm\to\KS\pi^\pm$}
}

%
\author{P.~del~Amo~Sanchez}
\author{J.~P.~Lees}
\author{V.~Poireau}
\author{E.~Prencipe}
\author{V.~Tisserand}
\affiliation{Laboratoire d'Annecy-le-Vieux de Physique des Particules (LAPP), Universit\'e de Savoie, CNRS/IN2P3,  F-74941 Annecy-Le-Vieux, France}
\author{J.~Garra~Tico}
\author{E.~Grauges}
\affiliation{Universitat de Barcelona, Facultat de Fisica, Departament ECM, E-08028 Barcelona, Spain }
\author{M.~Martinelli$^{ab}$}
\author{D.~A.~Milanes}
\author{A.~Palano$^{ab}$ }
\author{M.~Pappagallo$^{ab}$ }
\affiliation{INFN Sezione di Bari$^{a}$; Dipartimento di Fisica, Universit\`a di Bari$^{b}$, I-70126 Bari, Italy }
% ...
\collaboration{The \babar\ Collaboration}
\noaffiliation

\begin{abstract} 
We report on a search for \CP violation in the decay $D^\pm\to\KS\pi^\pm$ 
using a data set corresponding to an integrated luminosity of $469\,\invfb$ 
collected with the \babar detector at the \pep2\ asymmetric energy \epem
storage rings.
The \CP-violating decay rate asymmetry $A_{\CP}$ 
is determined to be 
$(-0.44 \pm 0.13 \stat \pm 0.10 \syst)\%$, consistent with zero at 2.7 $\sigma$ 
and with the standard model prediction of $(-0.332 \pm 0.006)\%$.
This is currently the most precise measurement of this parameter. 
\end{abstract}
 

\vspace{-0.2cm}
  
\maketitle    

In the standard model (SM), \CP violation (CPV)
arises from the complex phase of the 
Cabibbo-Kobayashi-Maskawa (CKM) quark-mixing 
matrix~\cite{Kobayashi:1973fr}.  
Measurements of the 
CPV asymmetries in 
the $K$ and $B$ meson systems are consistent with 
expectations based on the SM and, together with 
theoretical inputs, lead to the 
determination of the parameters of the CKM matrix.
CPV has not yet been observed in the  
charm sector, where the theoretical predictions based 
on the SM for CPV asymmetries are at the level of 
$10^{-3}$ or below~\cite{Buccella:1994nf}. 

In this Letter we present a search for CPV
in the decay $D^\pm\to\KS\pi^\pm$ by measuring the 
CPV parameter $A_{\CP}$ defined as:
\begin{equation}
A_{\CP}=\frac{\Gamma(D^+\to\KS\pi^+)-\Gamma(D^-\to\KS\pi^-)}
{\Gamma(D^+\to\KS\pi^+)+\Gamma(D^-\to\KS\pi^-)},
\end{equation}
where $\Gamma$ is the partial decay width for this decay. 
This decay mode has been chosen because of its clean experimental signature. 
Although direct \CP violation due to interference between
Cabibbo-allowed and doubly Cabibbo-suppressed amplitudes is 
predicted to be negligible within the SM~\cite{Lipkin:1999qz}, $\Kz-\Kzb$ mixing induces a 
time-integrated \CP violating asymmetry of $(-0.332\pm 0.006)\,\%$~\cite{Nakamura:2010zzi}.
Contributions from non-SM processes may reduce the value of the measured 
$A_{\CP}$ or enhance it up to the level of one percent~\cite{Bigi:1994aw,Lipkin:1999qz}.
Therefore, a significant deviation of the $A_{\CP}$ measurement from pure $\Kz-\Kzb$
mixing effects would be evidence for the presence of new physics beyond the SM.
Due to the smallness of the expected value,
this measurement requires a large data sample and precise
control of the systematic uncertainties.
Previous measurements of $A_{\CP}$ have been reported by the CLEO-c 
($(-0.6\pm 1.0 \stat \pm 0.3 \syst)\%$~\cite{:2007zt}) and
Belle collaborations ($(-0.71\pm 0.19 \stat \pm 0.20 \syst)\%$~\cite{Ko:2010ng}).

The data used in this analysis were recorded at or near the
$\Y4S$ resonance by the \babar detector at the \pep2 storage rings. 
The \babar detector is described in detail elsewhere~\cite{Aubert:2001tu}. 
The data sample corresponds to an integrated luminosity of $469\,\invfb$.
To avoid any bias from adapting the analysis procedure to the data, we 
perform a ``blind'' analysis where all aspects of the analysis, including 
the statistical and systematic uncertainties, are validated with data and 
Monte Carlo (MC) simulation based on GEANT4~\cite{Agostinelli:2002hh}
before looking at the value of $A_{\CP}$.
The MC samples include $\epem\to\qqbar\;(q=u,d,s,c)$ events, simulated with 
JETSET~\cite{Sjostrand:2006za} and \BB decays simulated with the EvtGen 
generator~\cite{Lange:2001uf}.
The coordinate system defined in \cite{Aubert:2001tu}
is assumed throughout the Letter.

We select $D^\pm\to\KS\pi^\pm$ decays by combining a $\KS$ 
candidate reconstructed in the decay mode
$\KS\to\pi^+\pi^-$ with a charged pion candidate.
A \KS candidate is reconstructed from two oppositely charged
tracks with an invariant mass within 
$\pm 10\,\mevcc$ of the nominal \KS mass~\cite{Nakamura:2010zzi},
which is equivalent to slightly more than $\pm 2.5\,\sigma$ in the measured
\KS mass resolution. The $\chi^2$ probability of 
the $\pi^+\pi^-$ vertex fit must be greater than $0.1\,\%$.
To reduce combinatorial background, we require the measured flight length
of the \KS candidate to be greater than 3 times its uncertainty. 
A reconstructed charged track that has $p_T\ge 400\,\mevc$ 
is selected as a pion candidate, where $p_T$ is the magnitude  
of the momentum in the plane perpendicular to the z axis.
At \babar, charged hadron identification is achieved through
measurements of ionization energy loss in the tracking system
and the Cherenkov angle obtained from a detector of internally 
reflected Cherenkov light. A CsI(Tl) electromagnetic calorimeter 
provides photon detection, electron identification, and neutral pion 
reconstruction~\cite{Aubert:2001tu}. 
In our measurement, the pion candidate is required not to be identified 
as a kaon, a proton, or an electron.
These selection criteria for the pion candidate are very effective in reducing
the charge asymmetry from track reconstruction and identification, as inferred from studying
the large control sample described later.
A kinematic vertex fit to the whole decay tree is then performed
with no additional constraints~\cite{Hulsbergen:2005pu}.
We retain only $D^\pm$ candidates having a $\chi^2$ probability for this fit
greater than 0.1\% and an invariant mass $m(\KS \pi^\pm)$ within $\pm65\mevcc$ of the 
nominal $D^+$ mass~\cite{Nakamura:2010zzi},
which is equivalent to more than $\pm 8\,\sigma$ in the measured
$D^\pm$ mass resolution.
Motivated by Monte Carlo simulation studies, 
we further require the magnitude of the $D^\pm$ candidate momentum in the 
$\epem$ center-of-mass (CM) system, $p^*(D^\pm)$, to be between $2$ and $5\,\gevc$.
This criterion reduces the combinatorial background to an acceptable level,
but also keeps some $D^\pm$ mesons from $B$ mesons decays (they are $\approx 8\%$
of the selected sample)\cite{CPTinv}.
Additional background rejection is obtained by requiring that the impact parameter 
of the $D^\pm$ candidate with respect to the beam-spot~\cite{Aubert:2001tu}, 
projected onto the plane perpendicular to the z axis,
be less than 0.3 cm and the $D^\pm$ lifetime $\tau_{xy}(D^\pm)$ be between $-12.5$ and $31.3$ ps. 
The lifetime is measured using $L_{xy}(D^\pm)$, defined as the distance of 
the $D^\pm$ decay vertex from the beam-spot
projected onto the plane perpendicular to the z axis.

To further improve the search sensitivity, a Boosted Decision Tree (BDT) 
algorithm~\cite{Speckmayer:2010zz} is constructed from seven 
discriminating variables for each $D^\pm$ candidate:
$\tau_{xy}(D^\pm)$, $L_{xy}(D^\pm)$,
the CM momentum magnitude $p^*(D^\pm)$, the momentum magnitudes and transverse components 
with respect to the beam axis for both the \KS and pion candidates.
Because all the input variables contains no charge information,
no charge bias is expected to be introduced by the algorithm
and this assumption has been verified using a large sample of MC simulated events.
The final selection criteria are based on the BDT output and optimized using 
truth-matched signal and background candidates from the MC sample.
For the optimization, we maximize the $S/\sqrt{S+B}$ ratio, 
where $S$ and $B$ are the numbers of signal and background
candidates whose invariant mass is within $\pm 31 \mevcc$ of the 
nominal $D^\pm$ mass.

\begin{figure}[tb]
\begin{center}
\includegraphics[width=0.45\textwidth]{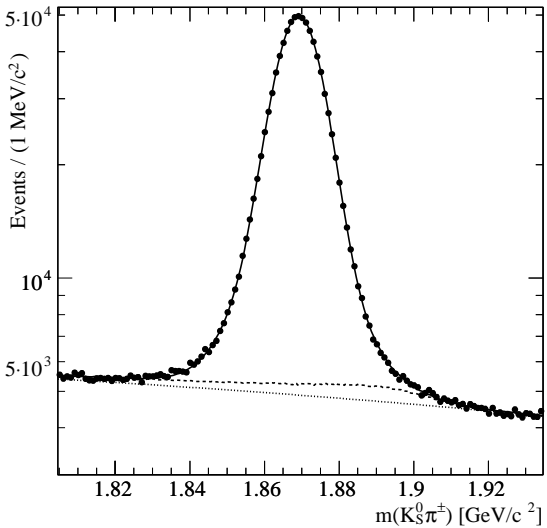}
\vspace{-0.3cm}
\caption{
Invariant mass distribution for $\KS \pi^\pm$ candidates 
in the data (black points).
The solid curve shows the fit to the data.
The dashed line is the sum of all backgrounds,
while the dotted line is combinatorial background only.
The vertical scale of the plot is logarithmic.}
\label{fig1}
\vspace{-0.7cm}
\end{center}
\end{figure}

A binned maximum likelihood (ML) fit to the $m(\KS \pi^\pm)$ distribution for the
retained $D^\pm$ candidates is used to extract the signal yield. 
The total probability density function (PDF) 
is the sum of signal and background components. The signal
PDF is modeled as a sum of three Gaussian functions, the first two of them with common mean.
The background PDF is taken as a sum of
two components: a background from 
$D^\pm_s\to\KS K^\pm$, where the $K^\pm$ is
misidentified as $\pi^\pm$, and a combinatorial background from other sources.
Based on MC studies, the yield of $D^\pm\to \pi^\pm\pi^\mp\pi^\pm$ decays in the
final data sample is estimated to be 0.02\% of the signal and the
estimated $A_{\CP}$ for this source to be less than 0.002\%.
Therefore a PDF to model this component is not included in the fit.
The background from the decay $D^\pm_s\to\KS K^\pm$ is modeled using a 
PDF sampled from the MC histogram for this mode. The combinatorial background is
described as a second-order polynomial.
The fit to the $m(\KS \pi^\pm)$ distribution yields 
$(807 \pm 1) \times 10^3$
signal events.
The data and the fit are shown in Fig.~\ref{fig1}.
All of the fit parameters are extracted from the fit to the data sample 
apart from the normalization of the background due to $D^\pm_s\to \KS K^\pm$, 
which is fixed to the value predicted by the MC simulation.

We determine $A_{\CP}$ by measuring the signal yield asymmetry $A$ defined as:
\begin{equation}
A=\frac{N_{D^+}-N_{D^-}}{N_{D^+}+N_{D^-}},
\end{equation}
where $N_{D^+}$($N_{D^-}$) is the number of fitted 
$D^+\to\KS\pi^+$($D^-\to\KS\pi^-$) decays.
The quantity $A$ is the result of two other contributions in addition to $A_{\CP}$. 
There is a physics component 
due to the forward-backward (FB) asymmetry ($A_{FB}$)
in $\epem\to\ccbar$, arising from $\gamma^*$-$Z^0$ interference and 
high order QED processes in $\epem \to \ccbar$. This asymmetry
will create a difference in the number of reconstructed $D^+$ and $D^-$ 
decays due to the FB detection asymmetries arising from the boost of the 
CM system relative to the laboratory frame.  
There is also a detector-induced component
due to the difference in the reconstruction efficiencies of $D^+\to 
K^0_s\pi^+$ and $D^-\to K^0_s\pi^-$ generated by
differences in the track reconstruction and identification efficiencies 
for $\pi^+$ and $\pi^-$. While $A_{FB}$ is measured together with $A_{\CP}$
using the selected dataset, we correct the dataset itself for the
reconstruction and identification effects using control data sets.

In this analysis we have developed a data-driven method to determine the 
charge asymmetry in track reconstruction as a function of the magnitude of 
the track momentum and its polar angle.
Since $B$ mesons are produced in the process $\epem\to\Y4S\to\BB$ nearly 
at rest in the CM frame and decay isotropically in the $B$ rest frame,
these events provide a very large control sample essentially free of 
any physics-induced charge asymmetry.
However, data recorded at the $\Y4S$ resonance also include continuum production
$\epem\to\qqbar\;(q=u,d,s,c)$, where there is a non-negligible FB asymmetry due to 
the interference between the single virtual photon process and other production processes,
as described above. 
The continuum contribution is estimated using the off-resonance data 
rescaled to the same luminosity as the on-resonance data sample.
Subtracting the number of reconstructed tracks in the rescaled 
off-resonance sample from the number of tracks in the on-resonance one, 
we obtain the number of tracks corresponding to the $B$ meson decays only.
Therefore, the relative detection and identification efficiencies of 
the positively and negatively charged particles for given selection criteria 
can be determined using the numbers of positively and negatively 
reconstructed tracks directly from data.

\begin{figure}[tb]
\begin{center}
\includegraphics[width=0.5\textwidth]{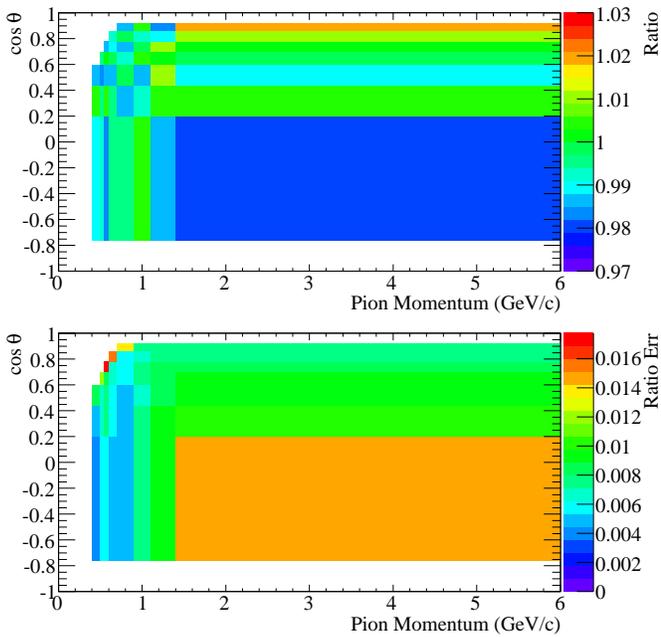}
\vspace{-0.3cm}
\caption{Map of the ratio between detection efficiency
for $\pi^+$ and $\pi^-$ (top) plus the
corresponding statistical errors (bottom). 
The map is produced using the numbers of $\pi^-$ and $\pi^+$ 
tracks in the selected control sample.}
\label{fig2}
\vspace{-0.7cm}
\end{center}
\end{figure}

Using samples of $8.5\,\invfb$ on-resonance and $9.5\,
\invfb$ off-resonance data,
applying the same charged pion track selection criteria used in the reconstruction 
of $D^\pm\to\KS\pi^\pm$ decays, 
and subtracting the off-resonance sample from the on-resonance sample,
we obtain a sample of more than 20 million tracks.
We use this sample to produce a map for the ratio of detection efficiencies
for $\pi^+$ and $\pi^-$ as a function of the track-momentum magnitude 
and $\cos\theta$, where $\theta$ is the polar angle of the track 
in the laboratory frame. The map and associated statistical errors
are shown in Fig.~\ref{fig2}.
Since the charm meson production is azimuthally uniform,
the $\phi$ dependence of this ratio is found to be very small and 
uncorrelated with momentum magnitude and polar angle.
Therefore, the ratio of detection efficiencies 
is averaged over the $\phi$ coordinate.
The statistical uncertainties can be reduced by increasing the control 
sample size, but this would bring a negligible reduction in the final 
systematic error.
In the fit procedure described below,
the $D^-$ yields, in intervals of pion-momentum and $\cos\theta$,
are weighted with this relative efficiency map to correct
for the detection efficiency differences between $\pi^+$ and $\pi^-$, 
leaving only FB and \CP asymmetries.
The average correction factor for each interval is $-0.09\%$.

Neglecting the second-order terms that contain the product of $A_{\CP}$ and $A_{FB}$,
the resulting asymmetry can be expressed simply as the sum of the two.
The parameter $A_{\CP}$ is independent of kinematic variables,
while $A_{FB}$ is an odd function of $\cos\theta^*_D$, 
where $\theta^*_D$ is the polar angle of the $D^\pm$ 
candidate momentum in the $\epem$ CM frame.
If we compute $A(+|\cos\theta^*_D|)$ for the $D^\pm$ candidates 
in a positive $\cos\theta^*_D$ bin and $A(-|\cos\theta^*_D|)$ for 
the candidates in its negative counterpart,
the contribution to the two asymmetries from $A_{\CP}$ is the same, 
while the contribution from $A_{FB}$ has the same magnitude but opposite sign.
Therefore $A_{\CP}$ and $A_{FB}$ can be written as a function of $|\cos\theta^*_D|$ as follows:
\begin{align}
  A_{FB}(|\cos\theta^*_D|) &= \frac{A(+|\cos\theta^*_D|) - A(-|\cos\theta^*_D|)}{2} \\ 
  \intertext{and}
  A_{\CP}(|\cos\theta^*_D|) &= \frac{A(+|\cos\theta^*_D|) + A(-|\cos\theta^*_D|)}{2}.
\label{eq:AcpAfb_intro}
\end{align}
Furthermore, the small fraction of the $D^\pm$ signal yields
produced from $B$ meson decays have zero FB asymmetry. 
As a result, the measured $A_{FB}$ from the 
$\epem\to\ccbar$ production is slightly diluted,
but the $A_{\CP}$ value is unaffected.

\begin{figure}[tb]
\begin{center}
\includegraphics[width=0.4\textwidth,clip=true]{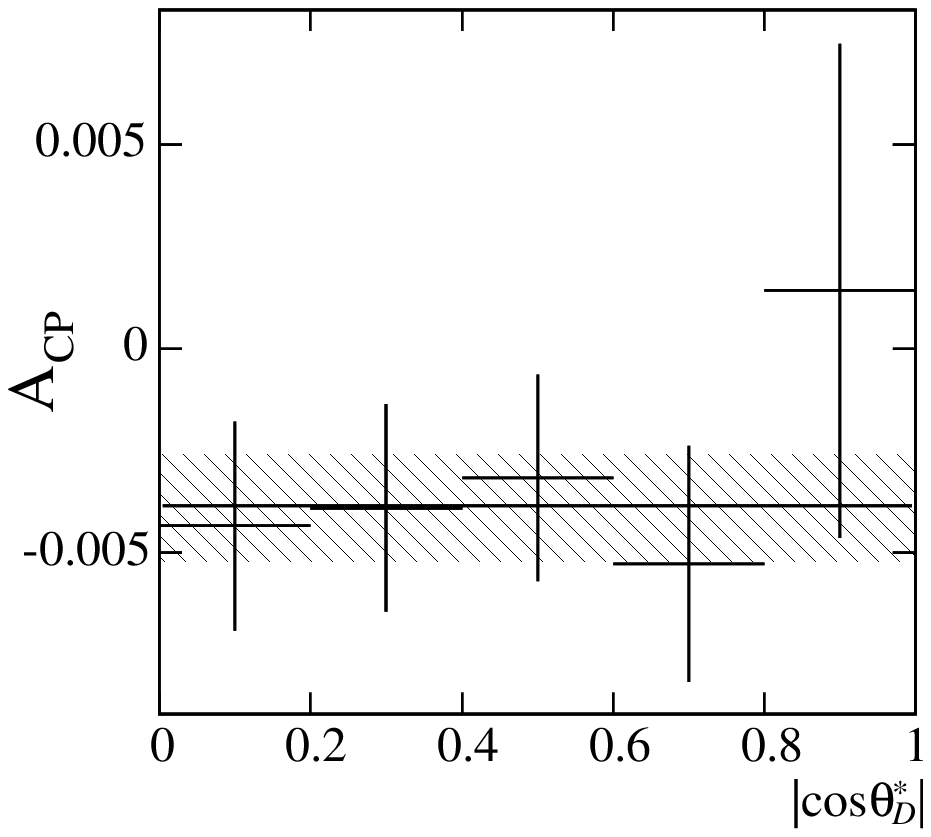}
\includegraphics[width=0.4\textwidth,clip=true]{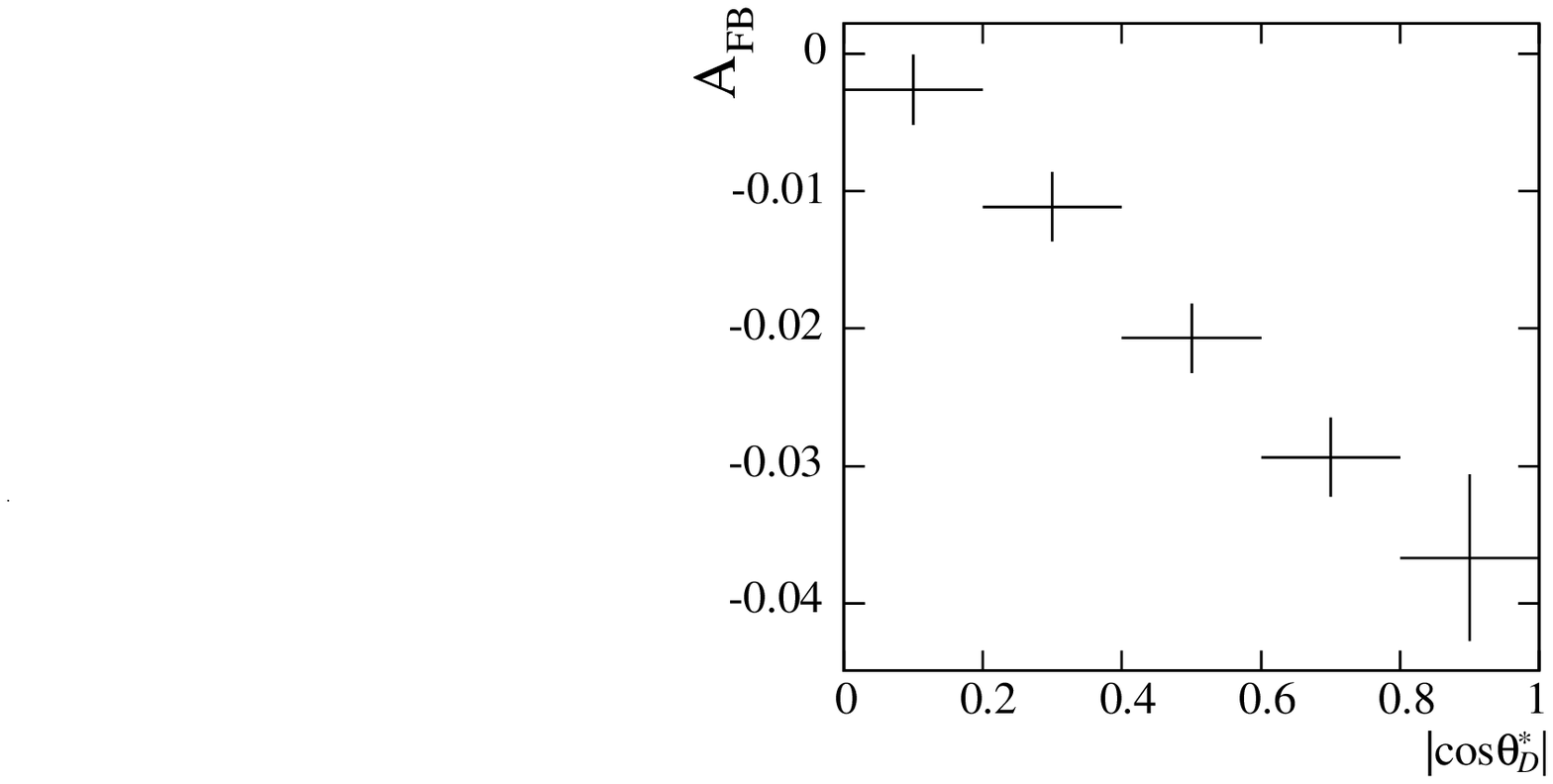}
\vspace{-0.3cm}
\caption{$A_{\CP}$ (top) and $A_{FB}$ (bottom)
asymmetries for $D^\pm\to\KS\pi^\pm$ candidates
as a function of $|\cos\theta^*_D|$ in the data sample.
The solid line represents the central value of
$A_{\CP}$ and the hatched region is the $\pm1\,\sigma$
interval, both obtained from a $\chi^2$ minimization
assuming no dependence on $|\cos\theta^*_D|$.}
\label{fig3}
\vspace{-0.7cm}
\end{center}
\end{figure}

The selected sample is divided into ten subsamples corresponding to
ten $\cos\theta^*_D$ bins of equal width and a simultaneous binned 
ML fit is performed on the invariant mass distributions of $D^+$ and 
$D^-$ candidates for each subsample to extract the signal yield asymmetries.
The PDF shape that describes the distribution in each subsample
is the same as that used in the fit to the full sample,
but the following parameters are allowed to float separately
in each subsample: the yields and the asymmetries for signal and combinatorial 
events, the mean of the second and third Gaussians for the signal PDF,
and the first order coefficient for the polynomial of the combinatorial background.
The relative fractions corresponding to the second Gaussian are allowed to float only 
for three high-statistics subsamples, while they have been fixed to zero for other ones
in order to have a converged fit.
The means of the three Gaussians for the signal PDF, the width of the first Gaussian, and the
second order coefficient for the polynomial of the combinatorial background
are allowed to float, but they have the same values for all the subsamples.
Therefore, the final fit involves a total of 78 free parameters.
Using the asymmetry measurements in five positive and in five negative 
$\cos\theta^*_D$ bins, we obtain five $A_{FB}$ and five $A_{CP}$ values. 
As $A_{CP}$ does not depend upon $\cos\theta^*_D$, we compute 
a central value of this parameter using a $\chi^2$ minimization
to a constant:
$A_{\CP}=(-0.39 \pm 0.13)\%$, where the error is statistical only.
The $A_{\CP}$ and $A_{FB}$ values are shown in Fig.~\ref{fig3}, together with
the central value and $\pm 1\,\sigma$ confidence interval for $A_{\CP}$.

We perform two tests to validate the analysis procedure. 
The first involves generating ensembles of toy MC
experiments and extracting $A_{\CP}$ for each experiment. We determine
that the fitted value of the $A_{\CP}$ parameter is unbiased, and that the fit
returns an accurate estimate of the statistical uncertainty. 
The second test involves fitting a large 
number of MC events from the full \babar detector simulation.
We measure $A_{\CP}$ from this MC sample to be within $\pm1\,\sigma$
from the generated value of zero.

The primary sources of systematic uncertainty are the contamination
in the composition of particles for the data control sample used to 
determine the charge asymmetry in track reconstruction efficiencies and 
statistical uncertainties in the detection efficiency ratios used to
weight the $D^-$ yields.
The charged pion sample selected to determine the
ratio of detection efficiencies for $\pi^-$ and $\pi^+$ 
contains a contamination of kaons, electrons,
muons, and protons at the percent level due to particle 
misidentification and inefficiencies. This contamination
introduces a small bias in the $A_{\CP}$ measurement due to
the slightly different particle identification efficiencies
between positively and negatively charged non-pion particles. 
The particle identification efficiencies,
measured in the data for positively and negatively charged tracks
using the method described in the previous paragraphs,
are found to be in a good agreement with the MC simulation. 
We therefore study this bias using the MC simulated 
events and determine the bias to be $+0.05\,\%$. 
As a result, we shift the measured $A_{\CP}$ by $-0.05\,\%$ to 
correct for the bias and then, conservatively, include the same value 
as a contribution to the systematic uncertainty.
Therefore the bias-corrected value of $A_{\CP}$ is $(-0.44 \pm 0.13)\%$.

The technique used here to remove the charge asymmetry 
from detector-induced effects produces a small systematic uncertainty in 
the measurement of $A_{CP}$ due to the statistical error in the relative efficiency map ($\pm0.06\%$).
Using MC simulation, we evaluate an additional systematic uncertainty
of $\pm0.01\%$ due to a possible charge asymmetry present in the control sample 
before applying the selection criteria.
Combining these two contributions with the systematic contribution from 
the difference in the composition of the control sample 
compared to the signal sample ($\pm0.05\%$), as described earlier, 
the total contribution from the correction technique is $\pm0.08\%$,
which is the dominant source of systematic error.
We also consider a possible systematic uncertainty due to the regeneration
of \Kz and \Kzb mesons in the material of the detector. 
\Kz and \Kzb mesons produced in the decay process can interact with the 
material around the interaction point before they decay.
Following a method similar to that described in~\cite{Ko:2010mk},
we compute the probability for \Kz and \Kzb to interact 
inside the \babar tracking system.
We numerically integrate the interaction probability distribution,
which depends on the measured nuclear cross-section 
for $K^\pm$ (assuming isospin symmetry),
the amount of material in the \babar\ beam-pipe and tracking detectors,
the \Kz/\Kzb time evolutions, 
and the \KS kinematic distribution and reconstruction efficiency
as determined from simulation studies.
From the difference between the interaction probabilities for \Kz and \Kzb,
we estimate a systematic uncertainty of $\pm0.06\%$.
Minor systematic uncertainties from the simultaneous ML fit are also considered:
the choice of the signal and background PDF, the limited MC data set 
to estimate the normalization of $D^\pm_s\to\KS K^\pm$, 
and the choice of binning in $\cos\theta^*_D$, for a total
contribution of $\pm0.01\%$.
The combined systematic uncertainty in the \CP asymmetry measurement 
including all the contributions is calculated as
the quadrature sum and is found to be $\pm0.10\%$.


In conclusion, we measure the direct \CP asymmetry, $A_{\CP}$, in the 
$D^\pm\to\KS\pi^\pm$ decay using approximately 800,000 $D^{\pm}$ signal candidates.
We obtain 
\begin{equation}
A_{\CP}=(-0.44\pm 0.13 \pm 0.10)\,\%,
\end{equation}
where the first error is statistical and the second is systematic. 
The result is consistent with the prediction of $(-0.332\pm0.006)\%$ 
for this mode based on the SM.

We are grateful for the excellent luminosity and machine conditions
provided by our \pep2 colleagues, 
and for the substantial dedicated effort from
the computing organizations that support \babar.
The collaborating institutions wish to thank 
SLAC for its support and kind hospitality. 
This work is supported by
DOE
and NSF (USA),
NSERC (Canada),
CEA and
CNRS-IN2P3
(France),
BMBF and DFG
(Germany),
INFN (Italy),
FOM (The Netherlands),
NFR (Norway),
MES (Russia),
MICIIN (Spain),
STFC (United Kingdom). 
Individuals have received support from the
Marie Curie EIF (European Union),
the A.~P.~Sloan Foundation (USA)
and the Binational Science Foundation (USA-Israel).


\begin{thebibliography}{99}

\bibitem{Kobayashi:1973fr}
\hyphenation{Ko-ba-ya-shi}
N.~Cabibbo, \prl {\bf 10}, 531 (1963); M.~Kobayashi and T.~Maskawa, \progtp {\bf 49}, 652 (1973).

\bibitem{Buccella:1994nf}
F.~Buccella {\it et al.},
Phys.\ Rev.\  D {\bf 51}, 3478 (1995).

\bibitem{Lipkin:1999qz}
H.~J.~Lipkin and Z.~Xing,
  Phys.\ Lett.\  B {\bf 450}, 405 (1999).

\bibitem{Nakamura:2010zzi}
  K.~Nakamura {\it et al.} (Particle Data Group),
  J.\ Phys.\ G {\bf 37}, 075021 (2010).

\bibitem{Bigi:1994aw}
  I.~I.~Bigi and H.~Yamamoto,
  Phys.\ Lett.\  B {\bf 349}, 363 (1995).

\bibitem{:2007zt}
  S.~Dobbs {\it et al.}  [CLEO Collaboration],
  Phys.\ Rev.\  D {\bf 76}, 112001 (2007)
  [arXiv:0709.3783 [hep-ex]].

\bibitem{Ko:2010ng}
  B.~R.~Ko {\it et al.}  (Belle collaboration),
  Phys.\ Rev.\ Lett.\  {\bf 104}, 181602 (2010).

\bibitem{Aubert:2001tu}
B.\ Aubert {\em et al.} (\babar\ Collaboration),
Nucl. Instr. Methods Phys. Res., Sect. A {\bf 479}, 1 (2002).

\bibitem{Hulsbergen:2005pu}
  W.~D.~Hulsbergen,
  Nucl.\ Instrum.\ Meth.\  A {\bf 552}, 566 (2005).

\bibitem{Agostinelli:2002hh}
S.~Agostinelli {\it et al.} (GEANT4 Collaboration),
Nucl. Instr. Methods Phys. Res., Sect. A {\bf 506}, 250 (2003).


\bibitem{Sjostrand:2006za}
  T.~Sjostrand, S.~Mrenna and P.~Z.~Skands,
  JHEP {\bf 0605}, 026 (2006)

\bibitem{Lange:2001uf}
  D.~J.~Lange,
  Nucl.\ Instrum.\ Meth.\  A {\bf 462}, 152 (2001).

\bibitem{CPTinv}
The contribution from \CP violation in $B$ decays from the Standard Model
processes is estimated to be negligible.

\bibitem{Speckmayer:2010zz}
  P.~Speckmayer, A.~Hocker, J.~Stelzer and H.~Voss,
  J.\ Phys.\ Conf.\ Ser.\  {\bf 219}, 032057 (2010).

\bibitem{Ko:2010mk}
B.~R.~Ko {\em et al.},
 arXiv:1006.1938 [hep-ex] (2010).

\end{thebibliography}
\end{document}